# Topological Edge States in Reconfigurable Multi-stable Mechanical Metamaterials

Zhen Wang[1,2,5], Feiyang Sun[3,5], Xiaodong Xu[3], Xin Li[4], Chuanqing Chen[2]*, Minghui Lu[2]*

[1] School of Electrical and Automation Engineering, Jiangsu Key Laboratory of 3D Printing Equipment and Manufacturing, Nanjing Normal University, Nanjing, China

[2] National Laboratory of Solid State Microstructures and Department of Materials Science and Engineering, Nanjing University, Nanjing 210093, China

[3] Department of Physics, MOE Key Laboratory of Modern Acoustics, Collaborative Innovation Center of Advanced Microstructures, Nanjing University, Nanjing 210093, China

[4] School of Mechanical Engineering, Nanjing University of Science and Technology, Nanjing, 210094, China

5 These authors contributed equally: Zhen Wang and Feiyang Sun

*Corresponding authors：Chuanqing Chen (chuanqingchen@nju.edu.cn), Minghui Lu (luminghui@nju.edu.cn)

**Abstract**:

Multi-stable mechanical structures find cutting-edge applications across various domains due to their reconfigurability, which offers innovative possibilities for engineering and technology advancements. This study explores the emergence of topological states in a one-dimensional chain-like multi-stable mechanical metamaterial composed of bistable units through a combination of mechanical and optical experiments. Drawing inspiration from the SSH (Su-Schrieffer-Heeger) model in condensed matter physics, we leverage the unique mechanical properties of the reconfigurable ligament-oscillator metamaterial to engineer a system with coexisting topological phases. Based on the one-dimensional periodic bistable chain, there is an exponential decay diffusion of elastic energy from both end boundaries towards the interior of the body. Experimental characterizations demonstrate the existence of stable topological phases within the reconfigurable multi-stable mechanical metamaterial. The findings underscore the potential of reconfigurable mechanical metamaterials as versatile platforms for flexibly exploring and manipulating topological phenomena, with applications ranging from impact resistance to energy harvesting and information processing.

**Keywords**: Multi-stable mechanical metamaterial, reconfigurability, SSH model, topological edge state

## 1. INTRODUCTION

Topological materials have revolutionized the field of condensed matter physics, offering novel properties and potential applications across various technological domains [1-9]. The discovery of topological insulators, which are characterized by robust conducting surface states protected by time-reversal symmetry, has sparked significant research endeavors in comprehending and leveraging topological phenomena. The observed topological non-triviality arises from the intrinsic properties of electronic Bloch states, rendering them applicable to periodic media of classical waves, including optical, acoustic, and mechanical lattices [10-18].

However, traditional single-stable materials lack controllability when studying topological states. Especially in verifying robustness, the need to artificially create defects imposes significant limitations on the practical applications of topological materials [19-21]. Fortunately, the realm of multi-stable mechanical metamaterials has burgeoned, affording the capacity to design and fabricate materials with tailored properties that are not inherent. Multi-stable mechanical metamaterials, typically achieved through series and/or parallel connections of artificially designed bistable units, showcase the ability to manifest numerous stable configurations under external loads, enabling reversible switching [22, 23]. The bistable unit traverses the energy barrier, swiftly transitions from the initial steady state to the second when the external load surpasses the critical load for steady-state transition. Importantly, the deformation state can be sustained without the need for continuous energy input.

In the exploration of topological states, a notable instance is the Su-Schrieffer-Heeger (SSH) model, originally conceived to elucidate electronic, photonic and phononic distribution in one-dimensional periodic chain [24-26]. The SSH model undergoes a topological phase transition characterized by the emergence of non-trivial edge states [27-29]. However, the investigation of the aforementioned characteristics faces challenges on two fronts. On one hand, this requires multiple sample preparations during the experimental process, which is time-consuming; on the other hand, the study of robustness entails the artificial introduction of defects, potentially limiting its practical applications. The primary obstacle lies in the significant challenges associated with achieving a reconfigurable SSH model.

Motivated by these advancements and the associated challenges, we delve into the prospect of inducing topological states in multi-stable mechanical metamaterials, akin to the SSH model. The mechanical universal testing machine is employed to assess the multi-stable characteristics of the structure. Utilizing continuous laser measurements of surface vibrations on the metamaterial, we substantiate the existence of stable topological phase within the metamaterial. This approach offers the advantages of non-contact and non-destructive detection. The existence of topological states in reconfigurable mechanical metamaterials hold significant implications for both fundamental physics and practical applications. Such systems provide a platform for exploring and manipulating topological phenomena in the realm of mechanical systems, complementing the advancements in electronic, photonic and phononic topological materials [30-33]. Moreover, the ability to achieve stable coexisting topological phases in metamaterials opens new avenues for the design of robust devices and applications, such as waveguiding, energy harvesting, and information processing.

This paper comprehensively investigates the topological phase in a reconfigurable multi-stable mechanical metamaterial through experimental and numerical analyses. We detail the unit cell structure, numerical simulations, and modeling results in Sections 2, 3 and 4, providing insights into the underlying mechanisms governing the metamaterial's topological behavior. Section 5 focuses on experimental characterization, and in the concluding section, we discuss broader implications and potential applications. The findings contribute to understanding topological mechanics and offer a basis for developing innovative mechanically engineered materials and devices.

**2. THEORY**

To understand the topological behavior in multi-stable mechanical metamaterials, the one-dimensional Su-Schrieffer-Heeger (SSH) model was introduced as a foundational theoretical framework. The SSH model describes the distribution of elastic energy through a periodic lattice of coupled mechanical oscillators. By deriving the Hamiltonian in matrix form, considering interatomic coupling strengths and phase differences between neighboring oscillators, the conditions for opening and closing the bandgap can be established. These conditions dictate the emergence of topological states in the mechanical metamaterials. Additionally, we investigate the role of parameters such as coupling constants between an individual structure and its left and right neighbors. ($t_1$, $t_2$) in tuning the bandgap and topological properties.

Our tight-binding model for the bistable lattice is an extension of the 1D Su–Schrieffer–Heeger model[34], with the Hamiltonian given as follows:

$$\mathrm{H} = n \sum (t_1 |n, A\rangle\langle n, B| + t_2 |n, B\rangle\langle n+1, A| + \mathrm{H.c.}) \tag{1}$$

Here, $|n, A\rangle$ represents the phonon state located at position A in the $n_{th}$ unit cell, while $|n, B\rangle$ corresponds to the phonon state at position B in the $n_{th}$ unit cell. The coefficients $t_1$ and $t_2$ denote the transition strengths within the lattice and between unit cells, with H.c. indicating the Hermitian conjugate term. This Hamiltonian characterizes the transition behavior of phonons in a 1D periodic lattice. The values of $t_1$ and $t_2$ can be tailored based on material properties, structural design, or experimental adjustments. This may involve selecting specific materials with desired mechanical characteristics, optimizing the structural layout, or fine-tuning parameters during experimental testing. By adjusting the numerical values of the transition strengths $t_1$ and $t_2$, different topological terms can be achieved.

And its bulk topology can be characterized by the polarization [35, 36], expressed as:

$$p = \frac{1}{2\pi}\int_{-\pi}^{\pi} \langle u_k | \frac{\partial}{\partial k} | u_k \rangle dk \qquad (2)$$

Here, $p$ represents the topological polarization, $\langle u_k |$ denotes the inner product of the energy band wave functions, and $\frac{\partial}{\partial k}$ indicates the partial derivative with respect to the wave vector $k$. This formula computes, within the Brillouin zone (BZ), for each wave vector $k$, the product of the inner product of wave functions and the wave vector derivative, integrating it over the entire Brillouin zone. This yields a topological invariant, namely the topological polarization. The result of the topological polarization is a quantized value, and in the presence of edge states, the topological polarization is non-zero. It serves as a characterization of the system's topological phase, such as in the cases of topological insulators and topological metals [37-39].

## 3. STRUCTURE DESIGN

This paper introduces a ligament-oscillator structure, which is originally derived from a one-dimensional bistable curved beam (the diagram depicted in fig. 1(a)). The curved beam is fully constrained on both sides. A horizontal straight beam of length $w$ (depicted in gray color) is positioned at the center of the curved beam, connecting to curved beams of length $L/2$ on both sides. These curved beams exhibit complete symmetry with respect to the straight beam. The incorporation of the horizontal section serves the purpose of minimizing rotational deformation during the transition between different steady-state deformations, thus significantly improving the overall stability of the structure. Under a Cartesian coordinate system, the configuration of the left-side curved beam in stable state 1 (depicted by a solid line) is defined by the equation $y = \frac{h}{2}\left[1 - \cos\frac{2\pi}{L}\right]$, where $h$ represents the beam height. Stable state 2 (depicted by a dashed line) is achieved when the applied load on the beam's apex surpasses a threshold value $F_c$. Moreover, the bi-stability of the curved beam is determined by the geometric constant of the ratio of the beam height to thickness $Q = h/t$. The constrained curved beam exhibits bi-stability when $Q > 2.31$ [40] and in this study, the $Q$ value is specified as 4.09.

Expanding upon the geometry concept of the bistable beam, a ligament-oscillator element including two symmetrical bistable beams distributed along the centerline and a quadrilateral star-shaped local resonance oscillator is designed as shown in fig. 1 (b). The key parameters of bistable ligament beam include its thickness ($t$=0.88 mm), apex height ($h$ =2.95mm), and span ($L + w$ =33.17mm) in the ligament-oscillator structure. The straight beam section is replaced by a block with identical height and width $l = w$ =7.37 mm to conveniently form a one-dimensional periodic chain of ligament-oscillator structure. The two bistable beams are interconnected through a centrally positioned quadrilateral star-shaped solid characterized by four arc sides with a radius of $R$=30.30mm, which is also functioned as a local resonator owing to its relatively large mass. The distance from the bottom of the bistable curved beam to the apex of the quadrilateral star-shaped arc side is denoted as $H$, where $H=7.37$mm$>h$, ensuring an ample deformation space for the bistable ligament beam to transition into stable state 2. The corners directly linked to the bistable beams within the quadrilateral star solid have a width of $T=2.21$mm, ensuring sufficient deformation stiffness to effectively substitute fully fixed boundary conditions on the bistable beam edges in the conceptual diagram. The remaining two sides are directly linked to the corners of the quadrilateral star-shaped solid through straight beams with a thickness $t=0.88$mm, and the separation distance between the outer faces of two straight beams is $b=39.36$mm. The ligament-oscillator element has a width ($a$) of 60mm and a depth ($D$) of 14.7mm.

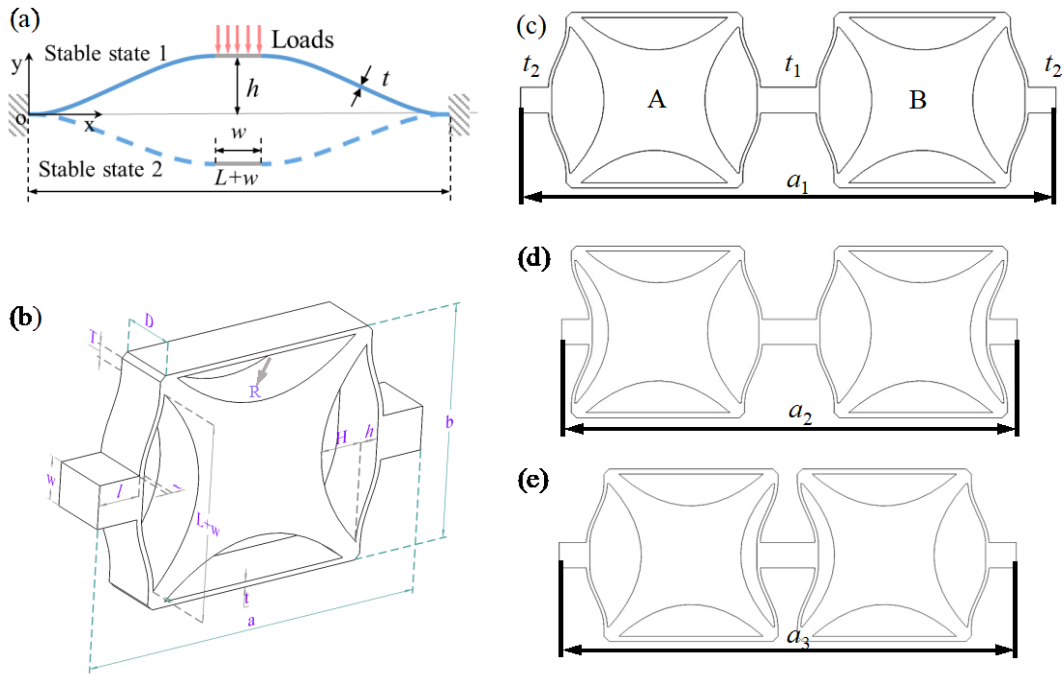

**FIG. 1.** (a) Schematic diagram of bistable curved beam, where solid and dashed lines represent stable state 1 and stable state 2. (b) Ligament-oscillator element with geometry parameters. Ligament-oscillator unit cell configurations with different

lattice constants: (c) $a_1$, (d) $a_2$, (e) $a_3$. A and B represent the two atoms within a unit cell, with $t1$ and $t2$ denoting the coupling strengths between them.

In alignment with the theoretical framework of the SSH model, a one-dimensional chain-like multi-stable metamaterial featuring a unit cell composed of two ligament-oscillator bistable elements was devised. The unit cell, including 4 bistable ligaments, theoretically harbors 16 distinct stable states due to the beams' bi-stabilities. Three specific stable states are chosen to symbolize varied intensity relationships within and between unit cell, as depicted in figs. 1 (c), (d), and (e). In the configuration illustrated in fig. 1(c), all four ligaments are in stable state 1, and its lattice constant is denoted as $a_1$, which equals $2a$. The distance of two adjacent bistable elements, whether within a unit cell or between adjacent unit cells, remains consistent, indicating a uniform transition strength relationship denoted as $t_1=t_2$. In the configuration illustrated in fig. 1(d), the central ligaments within the unit cell are in stable state 1, while the edge ligaments are in stable state 2, with its lattice constant $a_2$=105.02 mm, which is less than $a_1$. This means that the distance of adjacent ligament-oscillator elements within a unit cell is greater than that of adjacent units across different unit cells, implying $t_1>t_2$. While sharing the same lattice constant as the configuration in fig. 1(d) ($a_2=a_3$), the configuration depicted in fig. 1(e) presents a complete contrast: the central ligaments within the unit cell are in stable state 2, while the edge ligaments are in stable state 1. This indicates that the distance of adjacent ligament-oscillator elements within a unit cell is greater than that of adjacent units across different unit cells, implying $t_1<t_2$. To simplify the categorization of these three unit cell configurations, a coefficient $\Delta$ is introduced, defined as $\Delta = t_1/t_2$. Accordingly, the unit cell configurations in figs. 1(c)-(e) can be characterized by $\Delta = 1, \Delta > 1$ and $\Delta < 1$.

In this study, the specimens used for mechanical and optical experiments are manufactured via 3D printing method. Specifically, digitized geometries were segmented into horizontal layers using IdealMaker slicing software and then realized with a 3D printer (HUAFAST HS-334) utilizing the Fused Deposition Method (FDM). To minimize voids within the specimens, a 100% infill rate was applied. PolyMax polylactic acid (PLA) filaments were selected for their exceptional toughness, ensuring large deformation requirements with minimal damage during mechanical bistable deformation. The corresponding material properties, including density, elastic modulus, yield stress and Poisson's ratio, are 1.27 g/cm3，1.67 GPa，39.93 Mpa and 0.35, which have been detaily reported in Ref. [41].

To validate the reconfigurability of the structure, compression mechanical experiments were conducted to study the bistable characteristics of the ligament-oscillator element using the universal testing machine from Shimadzu Corporation, Japan. The tested specimen, having a thickness ($D$) of 10 mm, deviates from the previously indicated thickness of 14.7mm due to constraints in the fixture of the mechanical testing machine. Nonetheless, this discrepancy will not affect the confirmation of the structure's bistable properties. Given the symmetrical distribution of two bistable beams along the centerline, only one beam was tested, as both exhibited identical force-displacement curves. The horizontal block part of the tested beam and the middle position of the structure are clamped using upper and lower fixtures as shown in fig. 2. During test, the lower indenter is fixed, and the upper indenter compresses vertically at a velocity of 5mm/min. To ensure the reliability of the experimental results, the tests were conducted with two repetitions. Fig. 2 displays the force-displacement curves obtained during the compression process of the beam in the repeated experiments, which displays a high agreement. The contact force exhibited positive stiffness, increasing with displacement until reaching a critical indentation of approximately 0.85mm, where a snap-through phenomenon in the contact force occurred. Subsequently, the force decreased with an increase in displacement, signifying a transition from compression to tension deformation, and resulting in a negative stiffness of the beam. At around a displacement of 3.6mm, the contact force dropped to zero and continued to decrease with increasing displacement, eventually becoming negative. When the contact force is negative, the structure enters a stable locked state, and even with the removal of external forces, the structure cannot revert to its original shape. This bi-stability phenomenon, as indicated in Ref. [40], implies that the beam exhibits two stable states. At approximately a displacement of 7.0mm, the contact force resumes its ascent, marking a shift to positive contact stiffness.

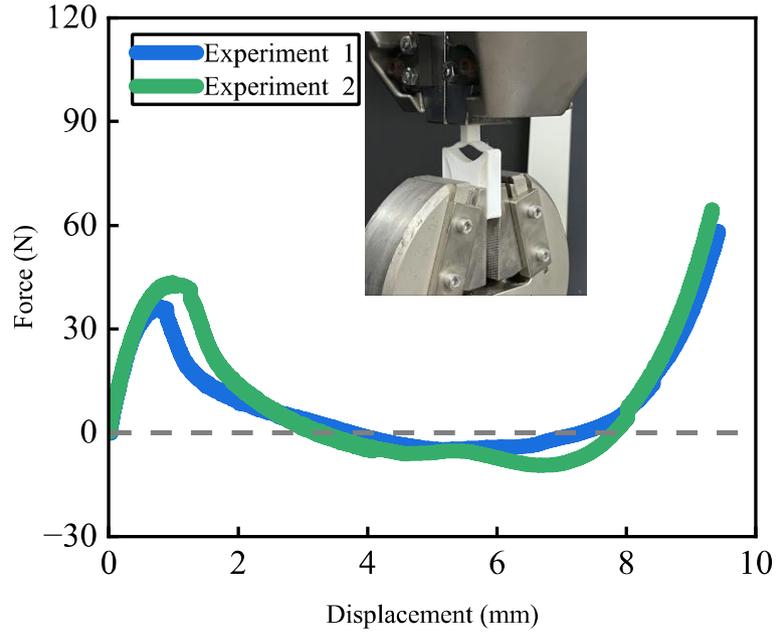

**FIG. 2.** Force-displacement curves for the cell unit under repeated compression. The green and blue lines show the force-displacement curves for two repeated compression experiments, respectively. The gray dotted line marks the boundary where the contact force is zero. The inset displays the specimen and the compression setup.

## 4. THE SIMULATIONS AND DISCUSSION

The ligament-oscillator diagram, when compared with the structure featuring a small filling ratio [42-44], exhibits significantly enhanced scattering and resonance. This characteristic proves advantageous for achieving localized edge states with a broader band gap, thereby ensuring high signal-to-noise performance. To assess the impact of varying transition strength on structural energy bands, comprehensive full-wave simulations are conducted using a commercial finite-element solver, specifically COMSOL Multiphysics, for the three lattice configurations illustrated in figs. 1(c)-(e). Figs. 3 (a)-(c) respectively illustrate the energy bands for the three unit cell configurations featured by $\Delta = 1, \Delta > 1$ and $\Delta < 1$. In the calculation, the Floquet periodicity boundary conditions are applied to the left and right surfaces of the block parts of unit cells to ensure the validity of the one-dimensional approximation. The other surfaces of the sample are set to free boundary conditions. As can be seen, for $\Delta = 1$, two pairs of pseudo-Kramer degeneracy points emerge between the first and second, and the third and fourth bands near the $k = \pm \pi$ point in the Brillouin zone. In this context, the blue energy bands correspond to in-plane modes, the black energy bands correspond to out-of-plane modes, and the red energy bands pass through the high-symmetry point $k = 0$, corresponding to the fundamental modes. When $\Delta > 1$, two pairs of non-degenerate points occur, resulting in the formation of two band gaps. This phenomenon is attributed to the breaking of translational symmetry, which originates from the different stable states of ligaments. The band gap for in-plane modes ranges from 33.82Hz to 40.29 Hz, while the band gap for out-of-plane modes corresponds to the range of 54.18Hz to 75.87Hz. When $\Delta < 1$, the energy bands are flipped on both sides of the band gap, although the width of the band gap remains the same as in the case of $\Delta > 1$.

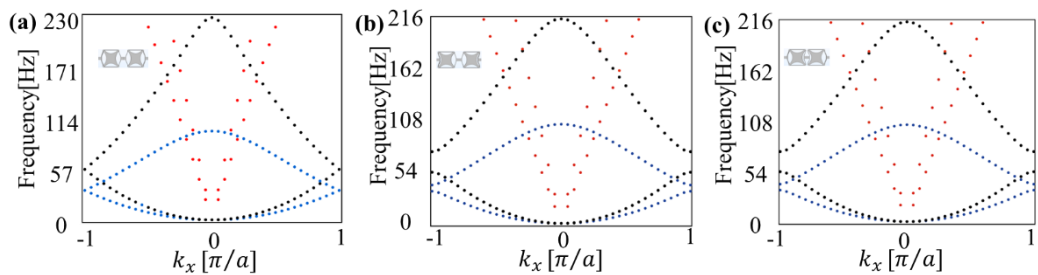

**FIG. 3.** Bulk energy bands for different unit cell configurations (the insets): (a) $\Delta = 1$, (b) $\Delta > 1$, and (c) $\Delta < 1$.

The out-of-plane mode with parabolic dispersion (indicated by a black dot) is primarily studied, particularly in non-contact optical measurements [12, 15]. The in-plane modes (represented by blue dots) exhibit weak coupling with the out-of-plane modes, resulting in only marginal effects in experiments, hence they can be safely neglected. This subsection focuses on the phase evolution of out-of-plane modes along the Brillouin zone boundary and the parity inversion of energy bands. Certainly, in the mechanical system, the parity of energy bands corresponds to the symmetry of modes. Figs. 4 (a) and (b) display the energy bands of out-of-plane modes of the unit cell configurations with $\Delta > 1$ and $\Delta < 1$, respectively. The different colored dots in figs.4(a) and 4(b) represent the field distribution of the modes at $k = 0$ or $k = \pi/a$ in figs. 4(c) and 4(d). To illustrate the phase evolution of out-of-plane modes along the Brillouin zone boundary, the unit cell field distribution for the corresponding modes was extracted, as shown in fig. 4(c). When $\Delta > 1$, the phase difference is zero of the modal field distributions corresponding to the high-symmetry points $k = \pi/a$ and $k = 0$, (maintaining symmetry along the x and y axes, respectively), while, the phase difference is $\pi$ when $\Delta < 1$ (the symmetry inversion along the y-axis).

In general, the inversion of energy bands on both sides of the band gap is accompanied by changes in the Brillouin vectors. Next, the evolution of the Brillouin vectors near the high-symmetry point $k = \pi/a$ corresponding to the inversion energy bands of configurations characterized by $\Delta > 1$ and $\Delta < 1$, was delved into, as illustrated in fig. 4(d). Black arrows represent Brillouin vectors, while white arrows depict the collective motion direction of Brillouin vectors. When $\Delta > 1$, the low-frequency mode exhibits antisymmetric about the y-axis. Simultaneously, the high-frequency mode is symmetric about the y-axis. In this case, it signifies a topologically trivial band gap, with the corresponding bulk topological polarization $P = 0$. When $\Delta < 1$, at the high-symmetry point $k = \pi$, the symmetry of the mode precisely undergoes inversion, accompanied by changes in the parity of the modes. It is crucial to emphasize that the dispersion of energy bands and the elastic field distribution of these modes in this context are both protected by the structural symmetry.

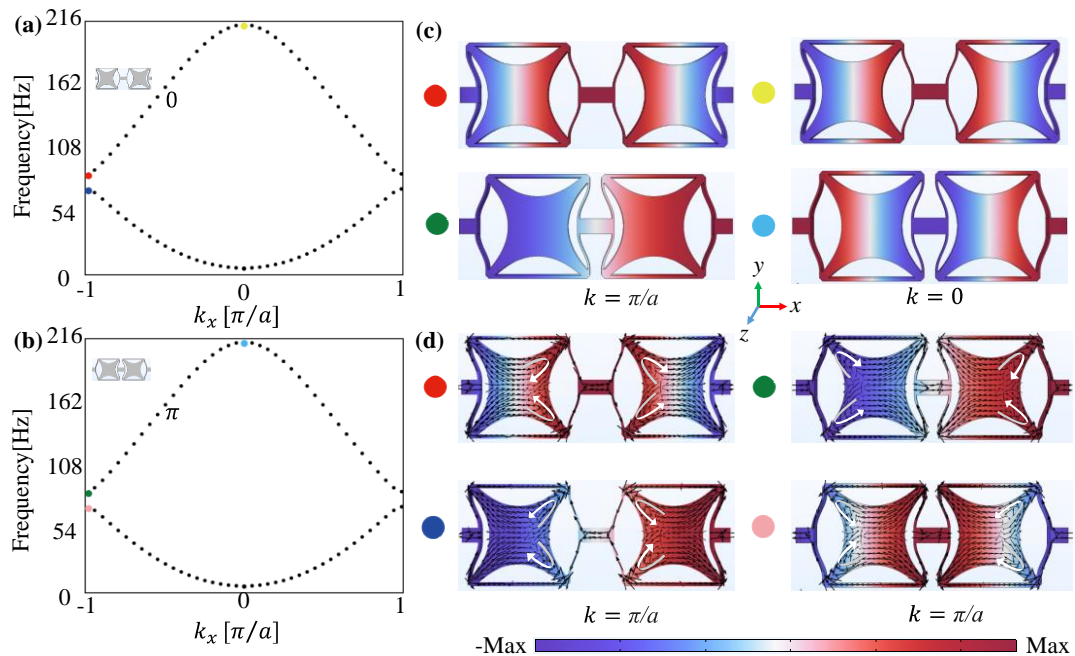

**FIG. 4.** The energy bands of out-of-plane modes correspond to: (a) $\Delta > 1$ and (b) $\Delta < 1$, respectively. (c) The corresponding elastic energy distribution of the bulk modes at the high-symmetry points (marked by different colors) $k = \pi/a$ and $k = 0$ in the Brillouin zone for $\Delta > 1$ and $\Delta < 1$, respectively. (d) The corresponding elastic energy distribution and Brillouin vector distribution of the energy band inversion of the bulk modes at the edge of the Brillouin zone.

Unlike omnidirectional band gaps in acoustic and elastic systems [45-49], the out-of-plane band gap studied in this paper includes two fundamental bulk modes, as illustrated by the red energy bands in fig. 5(a). Therefore, to explore the topological edge states within the band gap, it is necessary to characterize and exclude the characteristics of the fundamental bulk modes within the band gap, simultaneously eliminating the influence of the fundamental bulk modes on the topological edge states. The mode responses corresponding to these two fundamental bulk modes at $k = 0.2\,\pi/a$ were explored, as depicted in fig. 5(b). The elastic energy distribution of the high-frequency mode exhibits symmetry about both the *x* and *y* axes, whereas the distribution of the low-frequency mode lacks symmetry. It is noteworthy that the elastic energy of both two bands is primarily concentrated on the central oscillator and ligament, with weak energy distribution at the connections between unit cells. This phenomenon differs significantly from the out-of-plane modes depicted in fig. 4, which exhibits strong elastic energy distribution at the connections between unit cells. This observation provides valuable insights for subsequent analysis of bulk modes within the band gap in figs. 6 and 7.

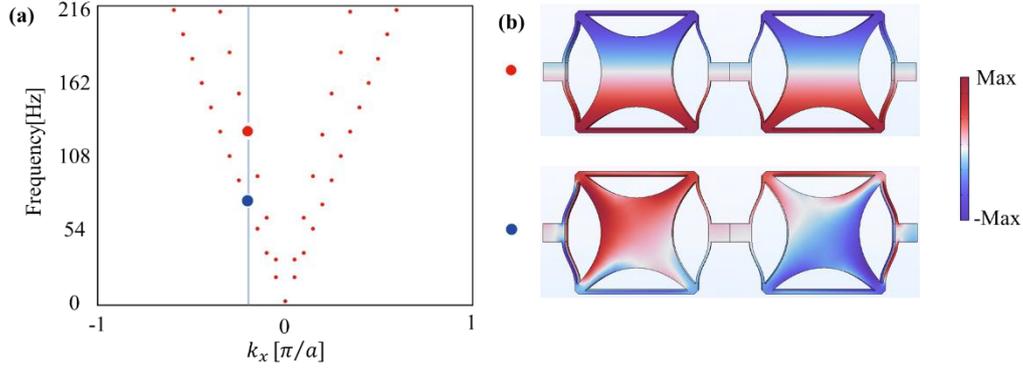

**FIG. 5.** The two fundamental bulk modes intersect at $k = 0.2\,\pi/a$. (a) Two zero-dispersion fundamental modes, symmetric about $k = 0$. (b) Field distribution of the unit cell when $k = 0.2\,\pi/a$, corresponding to the intersections of the sky-blue vertical line with the bands in (a).

Topological edge states, specifically those with non-zero topological polarization $(P \neq 0)$, have garnered widespread attention due to their defect-immune characteristics [50-52]. Indeed, leveraging the reconfigurability of multi-stable metamaterials, the distribution of the elastic field in periodically structured materials with different $\Delta$ values can be investigated by controlling the structure's distinct stable states. Through the periodic distribution of unit cell configurations characterized by $\Delta > 1$ and $\Delta < 1$, as shown in figs. 1(d) and (e), two types of one-dimensional chain-like multi-stable metamaterials can be obtained. Numerical simulations, utilizing a total of 8 unit cells, were conducted to calculate the eigenfrequencies of these metamaterials. The results are illustrated in figs. 6 (a) and (c), with the horizontal axis representing the mode number and the vertical axis displaying 54 eigenfrequencies. When $\Delta > 1$, the band gap is topologically trivial, with its range covered by a light blue transparent shadow from 54.18Hz to 75.87Hz.

Bulk modes corresponding to the dark blue dots within the bandgap frequencies of 60.55 Hz and 70.47 Hz are illustrated in fig. 6(b), respectively. These bulk modes coexist with the topological edge states within the band gap. When attempting to excite edge states within the band gap, the influence of bulk modes is likely, especially if their frequencies are close. When $\Delta<1$, the band gap is topologically nontrivial. The out-of-plane elastic field distribution of modes within the band gap, with bulk frequencies of 59.63 Hz, and 69.69 Hz (dark blue dots), as well as edge mode frequencies of 60.82 Hz and 66.44 Hz (red dots), are depicted in fig. 6(d), respectively. Compared to the condition of $\Delta>1$, in addition to the two dark blue bulk modes, there are two degenerate edge modes with elastic energy mainly localized at the connections between unit cells and the ligament.

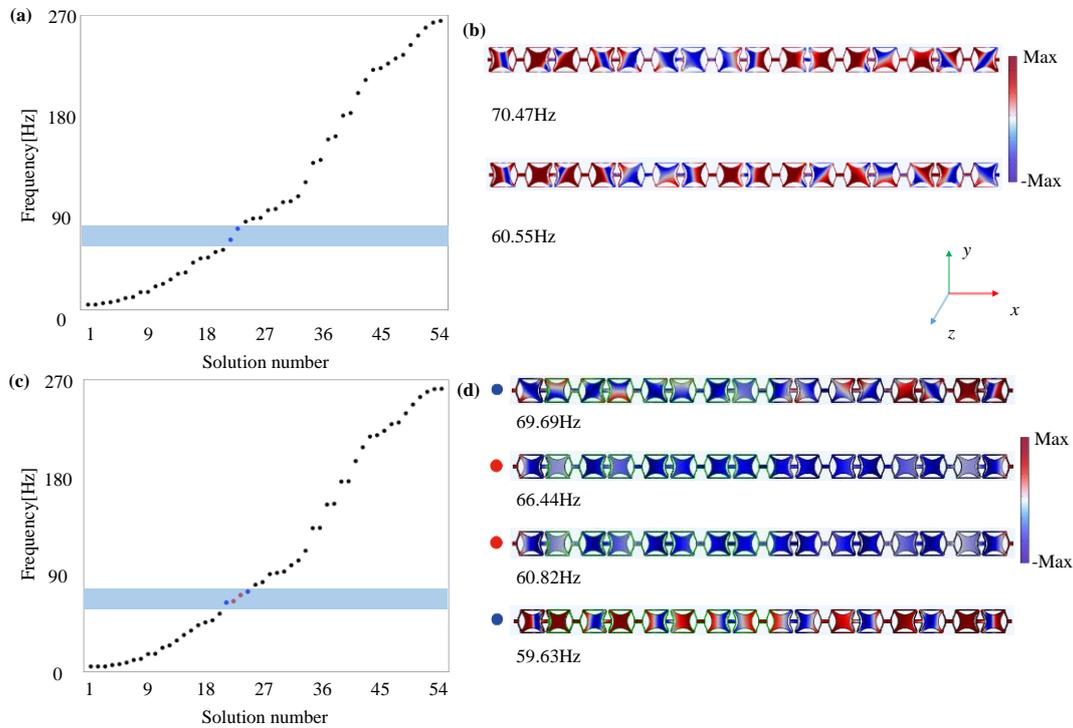

**FIG. 6.** The eigenmodes of the one-dimensional periodic chain when (a) $\Delta>1$ and (c) $\Delta<1$ with the horizontal axis representing mode numbers (54) and the vertical

axis indicating the corresponding eigenfrequencies. The light blue shadow denotes the bandgap range. (b) Bulk modes corresponding to the dark blue dots within the bandgap, with frequencies of 60.55 Hz and 70.47 Hz when $\Delta > 1$. (d) Bulk and edge modes corresponding to the dark blue and red dots within the bandgap, with bulk frequencies of 59.63 Hz, and 69.69 Hz, and edge mode frequencies of 60.82 Hz and 66.44 Hz when $\Delta < 1$. The black dots outside the band gap represent bulk modes in (a) and (c).

To verify the chiral symmetry of the edge states, the one-dimensional periodic chain-like multi-stable metamaterial with $\Delta < 1$ is truncated at the right edge to break its mirror symmetry. Hence, the chain structure consists of 7 unit cells and an additional bistable element, with the left ligament in stable state 1 and the right ligament in stable state 2. The corresponding calculations of the eigenfrequencies are illustrated in fig. 7 (a), where the shaded region corresponds to the bulk band gap. Comparing fig. 6 (c), the disappearance of the two degenerate edge modes is notable, replaced by a chiral state localized at the left edge. This change is attributed to the absence of mirror symmetry. Additionally, within the band gap, two bulk modes, represented by dark blue dots, are present. The corresponding out-of-plane elastic field for these bulk modes is depicted in fig. 7(b). The bulk modes within the band gap in figs. 6 and 7 correspond to the two zero-dispersion fundamental bulk modes in fig. 5. The distribution of elastic energy is primarily concentrated on the central mass block of the bistable structure. This energy distribution differs from that of the topological edge states, with the elastic energy mainly localized on the ligaments.

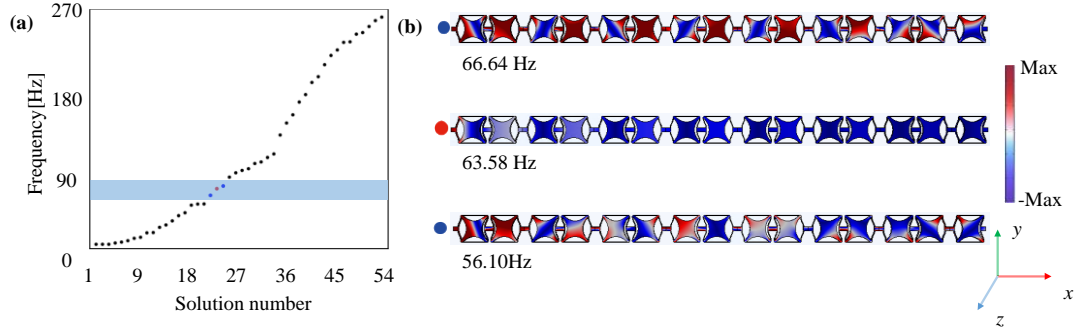

**FIG. 7.** Eigenmodes of the one-dimensional periodic bistable chain when mirror symmetry is broken ($\Delta < 1$). (a) Distribution plot of eigenmodes corresponding to a lattice periodicity of 7+1/2 unit cells, with the horizontal axis representing mode numbers (54) and the vertical axis indicating the corresponding eigenfrequencies. The light blue shadow denotes the bandgap range. (b) Bulk and edge modes corresponding to the dark blue and red dots within the bandgap, with bulk frequencies of 56.10 Hz and 66.64 Hz, and edge mode frequencies of 63.58 Hz, respectively. The black dots outside the band gap represent bulk modes in (a).

## 5. EXPERIMENTS AND RESULTS

To validate the numerical simulation results, optical experiments were conducted using a continuous laser probe to detect the out-of-plane elastic energy at each lattice point of the one-dimensional periodic chain. Fig. 8 presents the sketch of the overall experimental set-up in (a), and the layout of the experimental site in (c) together with the details of the tested specimen in (b) and (d). As can be seen, the whole experiment system consists of a laser vibrometer, a modal exciter, a waveform generator, an oscilloscope, a computer and a test specimen. The chain-like multi-stable metamaterial with 8 unit cells was fabricated by composing a series of bistable elements as the tested specimen. Notably, multi-stable mechanical metamaterials, equipped with a tunable topological structure, enable the exploration of various topological configurations within a single specimen. Hence, structures featuring $\Delta = 1$ and $\Delta < 1$ are accomplished by tunning the stable states of ligaments in the same specimen, as shown in fig. 8 (d). Specifically, the specimen is suspended in mid-air using four threads to minimize structural deformation induced by gravity, thus approximating stress-free boundary conditions. The modal exciter SA-JZ002, with a bandwidth of 15 kHz, utilized a chirp signal (ranging from 0 to 100 Hz) generated by an Agilent arbitrary waveform generator as the sound source. Concurrently, the POLYTEC laser vibrometer OFV-5000 with the OFV-505 probe, boasting a bandwidth of 2.5 MHz, was employed to measure the out-of-plane velocity and displacement of the specimen. All measuring locations are represented by red points in the structural diagram, as shown in fig. 8 (b). Reflective tape applied to the sample surface during measurements enhanced the reflection signal. Throughout the experiment, the modal exciter was meticulously attached to the back surface of the specimen using epoxy acrylic adhesive, ensuring precise alignment with the laser focal points on the front surface of the specimen. The experimental test data are displayed in real time by an oscilloscope and finally stored. In the optical experiment, only the specimen with $\Delta < 1$ configuration was tested due to its topologically nontrivial band gap identified in the numerical analysis.

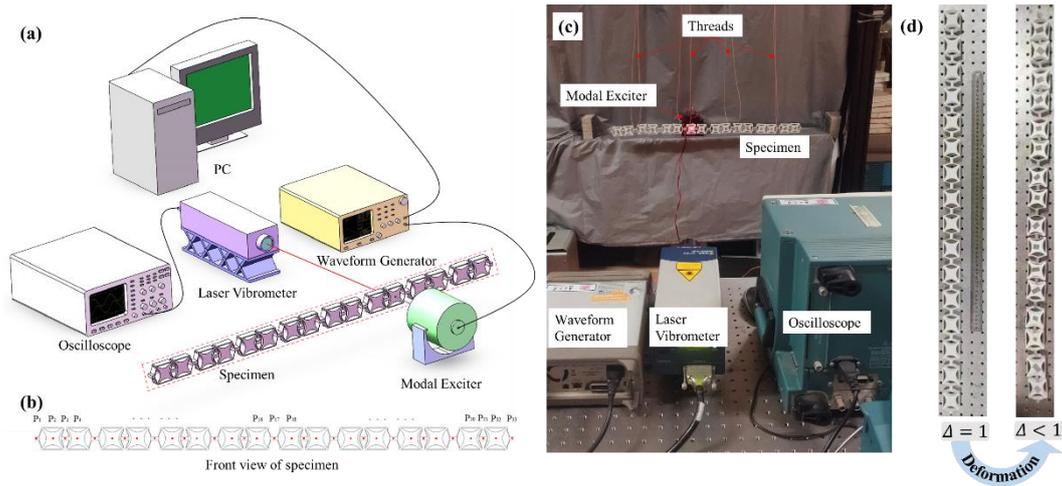

**FIG. 8.** (a) Schematic of the optical experiment system, including the laser vibrometer, a modal exciter, a waveform generator, an oscilloscope, a computer and a test specimen. (b) Experimental setup. (c) The front view of specimen diagram, with the red dot indicating the measured locations. (b) A same 3D printed specimen with different unit cell configurations ($\Delta = 1$ and $\Delta < 1$), achieved by tuning the stable states of the ligaments.

Fig. 9 presents a comparison between experimental results and numerical simulations, showcasing the normalized displacement amplitude (with respect to the amplitude of the edge state at the right boundary of the specimen) plotted against the measured positions (corresponding to the red dots in fig 8. (b)). Inevitably, due to the non-uniform distribution of stress during the deformation process, the specimen undergoes distortion, introducing random defects at the edges and within the body. The red dash-dot line in fig. 9 (a) corresponds to the numerically calculated results, representing the red edge state in fig. 5 (frequency 60.82 Hz), while the blue solid line represents the experimental measurement results (frequency approximately 34 Hz), with a frequency deviation of approximately 44%. The red dash-dot line in fig. 9 (b) corresponds to another red edge state in fig. 5 (frequency: 66.44 Hz), while the blue solid line represents the experimental measurement result (frequency approximately 36 Hz), with a frequency deviation of approximately 46%.

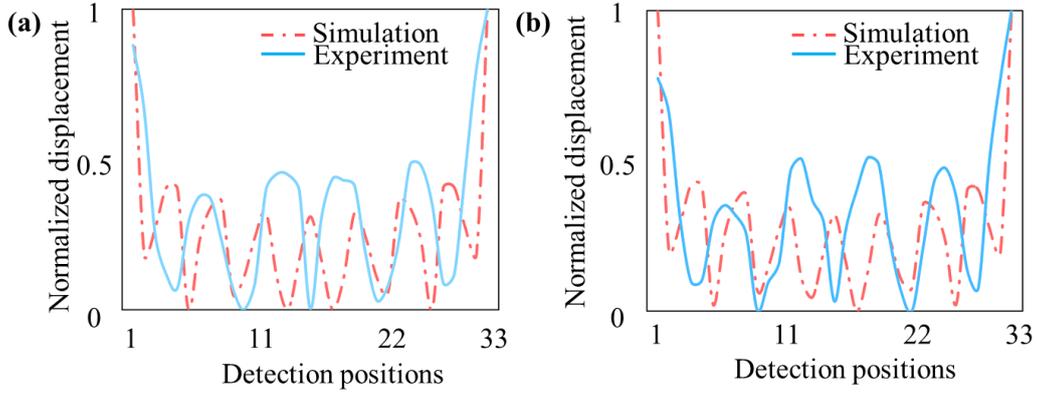

**FIG. 9.** The comparison of the normalized displacement amplitude vs. measured positions between experimental results and numerical simulations.

Of course, it is necessary to choose materials with good toughness for 3D printing to ensure that ligament fracture does not occur during modulation between the two stable states. On the other hand, during the modulation of the bistable states, due to uneven force, inevitable introduction of defects occurs. These random defects not only validate the robustness of topological edge states but also save resources compared to artificially introducing defects, such as cavities, disorders, or cracks, as done in other related works[14, 53, 54]. Therefore, these random defects can offer more compelling evidence for the robustness of the edge states.

The order of magnitude of the edge states and bulk states in the experimental results is consistent with the numerical calculations. Obviously, the left edge state shows a significant deviation compared to the right edge. Additionally, there are certain errors in the bulk states, along with a shift in peak values. The main reason is the breaking of the structural mirror symmetry due to the uneven distribution of stress during the transition between the two stable states. Furthermore, uneven stress loading results in an actual lattice constant that is slightly larger than the theoretically calculated value. As a result of these two effects, the frequencies of the edge states are shifted. If improvements are to be made in this experiment, it is necessary to search for materials with better toughness to ensure that the mirror symmetry is not compromised during the deformation process. Simultaneously, it is essential to ensure that the stress loading during the bistable transition deformation process remains as uniform as possible. In addition, the experimental results also were constrained by practical experimental conditions, such as the uneven bonding strength between the exciter and each data collection point on the sample surface, and the suspension of the sample, which cannot fully guarantee an ideal stress-free boundary.

## 6. SUMMARY AND CONCLUSIONS

Combining theory and experiments, the study observes a topological phase transition in the bistable lattice, revealing a band gap width of approximately 54.18-75.87 Hz. In the experiments, the gap shift rate is approximately 45%, attributed to the application of non-uniform loading. Analysis of energy flow and phase in the ligament-oscillator diagram, along with examinations of mirror symmetry influence, enhances the understanding of topological phases in mechanically bistable systems. Advanced optical instrumentation probes the elastic energy distribution of topological edge and bulk states within the one-dimensional periodic bistable lattice.

Experimental results closely align with theoretical calculations, validating proposed modes and predictions. Deeper analysis explores the impact of uneven loading during the experimental process on both edge and bulk states. In this work, through the ingenious structural design of a multi-stable metamaterial, we simultaneously investigate two distinct topological phases within the same sample. This significantly enhances the flexibility of experiments and substantially reduces experimental costs. Experimentally, even under significant deformations and approximately 45% shift in the band gap of the sample, the topological edge states exhibit remarkable robustness. This validates the necessity of studying topological edge states in multi-stable systems and underscores the reliable application prospects, such as novel low-frequency mechanical switches, structural health monitoring, and adaptive energy absorption systems, among others. Additionally, this study provides insights for designing versatile materials with enhanced functionalities, seamlessly integrating topological protection with mechanical tunability.

**Acknowledgments**

This work was supported by the National Key R&D Program of China (Grant No. 2023YFA1406904), the National Natural Science Foundation of China (Grant No. 52250363), the China Postdoctoral Science Function (Grant No. 2023M741612), the Jiangsu Funding Program for Excellent Postdoctoral Talent (Grant No. 2023ZB477), Postdoctoral Fellowship Program of CPSF (Grant No. GZC20231019), and 2022 Research Project on Basic Sciences (Natural Sciences) in Higher Education Institutions of Jiangsu Province.